\documentclass[twocolumn,showpacs,preprintnumbers,amsmath,amssymb]{revtex4}
\usepackage{graphicx}
\usepackage{dcolumn}
\usepackage{bm}

\begin{document}

\title{One Dimensional Magnetized TG Gas Properties in an External Magnetic Field}

\author{Zhao Liang Wang}
 \email{wzlcxl@mail.ustc.edu.cn}
\author{An Min Wang}
 \email{anmwang@ustc.edu.cn}
\affiliation{Department of Modern Physics, University of Science and
Technology of China, Hefei, Anhui, China.}

\begin{abstract}
With Girardeau's Fermi-Bose mapping, we have constructed the
eigenstates of a TG gas in an external magnetic field. When the
number of bosons $N$ is commensurate with the number of potential
cycles $M$, the probability of this TG gas in the ground state is bigger than the TG gas raised by
Girardeau in 1960. Through the comparison of properties between this TG
gas and Fermi gas, we find that the following issues are
always of the same: their average value of particle's
coordinate and potential energy, system's total momentum,
single-particle density and the pair distribution function. But the reduced single-particle matrices and their
momentum distributions between them are different.
\end{abstract}

\maketitle

\section{Introduction}\label{sec:level1}

The problem of a one-dimensional (1D) hard-core gas was first
studied classically by Tonks \cite{lab1}. Girardeau \cite{lab2,lab3}
continued the research in quantum, and put forward the Fermi-Bose
mapping to solve energy spectrum and the corresponding wave function
of a one dimensional Tonks-Girardeau (TG) gas. After that, Lieb and
Liniger \cite{lab4} considered $N$ Bose particles interacting via a
repulsive $\delta$-function potential with a coupling constant
$\gamma$. In fact, TG gas can be considered as the
Lieb-Liniger gas when $\gamma\rightarrow\infty$.

During the past decade, the study on TG gas has undergone a rapid development
in theoretical and experimental
\cite{lab21,lab5,lab6,lab7,lab8,lab9,lab10,lab20,lab11,lab12,lab13,lab14,lab15}.
In particular, A. Lenard \cite{lab21} indicated that the
off-diagonal parts of the one-body density matrix and the momentum
distribution show distinct differences in both the TG gas and the
Fermi gas. G. J. Lapeyre et al. \cite{lab7} studied the momentum
distribution of a harmonically trapped gas. The most important
progress is the realization of the strongly interacting bosons
\cite{lab14} in a one-dimensional optical-lattice trap and the TG
gas \cite{lab15} by trapping ${}^{87}$Rb atoms by trapping them with
a combination of two light traps.

We have known some properties of one dimensional TG gas, and we are interested in its properties and features in external magnetic field. Consequently, we first reveal some properties of both
gases in the same external magnetic field and compared them. With the Fermi-Bose mapping \cite{lab2,lab3}, we obtained all the eigenenergies and eigenstates of this TG gas. When the number of
bosons $N$ is commensurate with the number of potential cycles $M$,
the TG system in an external magnetic field will be in the ground
state with a bigger probability than the TG gas raised by Girardeau
in 1960. Many properties of the TG gas and the Fermi gas are always
of the same even if it's time-dependent, such as their average value
of particle's coordinate and potential energy, system's total
momentum, single-particle density and the pair distribution
function. But both their reduced single-particle matrices and momentum
distributions are different.

\section{TG gas in an external magnetic field}\label{sec:level2}

In a TG gas, the boson is assumed to have an ``impenetrable'' hard core
characterized by a radius of $a$. From Girardeau's work, with the
hard core radius $a\rightarrow0$, the interparticle interaction is
given by
\begin{equation}
U(x_{_i},x_{_j})=\begin{cases}
0, & \text{$x_{_i} \neq x_{_j}$},\\
\infty, &\text{$x_{_i}=x_{_j}$}.
\end{cases}
\end{equation}

Such an interparticle interaction could be represented by the
following subsidiary condition on the wave function $\psi$ :
\begin{equation}\label{eqn2}
\psi(x_{_1},\cdots,x_{_N},t)=0\,\,\,\mathrm{if}\,\,\,\hspace{\stretch{1}}x_{_i}=x_{_j},\hspace{\stretch{0.3}}1\leqslant
i<j\leqslant N.
\end{equation}

In order to let the TG gas wave function, which must be symmetric with respect to the permutations of any  $x_i$ and $x_j$, satisfy this condition (\ref{eqn2}), we need the ideas of Fermi-Bose mapping proposed by Girardeau \cite{lab2,lab3}. That is, using the fact that the Fermi wave function, denoted by $\psi^F$, satisfies condition (\ref{eqn2}) naturally and is antisymmetric, a bosonic wave function $\psi^B$ of a TG gas can be constructed by
\begin{equation}\label{eqn7}
\psi^B(x_1,\cdots,x_N,t)=A(x_1,\cdots,x_N)\psi^F(x_1,\cdots,x_N,t)
\end{equation}
in which
\begin{equation}
A(x_1,\cdots,x_N)\equiv \prod^N_{i>j}\mathrm{sgn}(x_i-x_j),
\end{equation}
and
\begin{equation*}
\mathrm{sgn}(x)\equiv \frac{x}{|x|}=\begin{cases} 1, & \text{$x>0$},\\ -1,
&\text{$x<0$}.
\end{cases}
\end{equation*}
As the ground state of a bose system is non-negative
\cite{lab16}, mapping (\ref{eqn7}) for the stationary ground state
of a system reduces to a simplified form \cite{lab2}
\begin{equation}\label{eqn8}
\psi^B(x_1,\cdots,x_N)=|\psi^F(x_1,\cdots,x_N)|
\end{equation}

To illustrate our general ideas, we now study the case that magnetized bosons in an external
magnetic field $B(x)=-B\cos(2\omega x)$. For simplicity, we take
$B(x)$ independent of $t$. therefore,
\begin{equation}\label{eqn9}
\hat{H}=\sum^N_{i=1} \left[-\frac{\hbar^2}{2m}\frac{\partial^2}{\partial
x^2_i}+V(x_i) \right].
\end{equation}
in which
\begin{equation}\label{eqn10}
V(x)=\begin{cases}\mu B\cos(2\omega x),&\text{$0\leqslant
x_i\leqslant \frac{L}{\omega}$}, \\ 0,&\text{else}.
\end{cases}
\end{equation}

We suppose $L=M\pi$, where $M$ is an integer and $\mu$ is the
magnetic moment of atom. For the sake of simplicity, we assume
that both $N$ and $M$ are odd (As a result of the introduction of
$A(x_{_1},\cdots,x_{_N})$, $\psi^B$ is periodic if $N$ is odd and
antiperiodic if $N$ is even \cite{lab2,lab20}). From Bloch theory,
the corresponding energy spectrum of system in periodic potentials
has the energy band structure. We use the notation $n$ as the band
index and $k$ as the Bloch wave vector in the first Brillouin zone.
For convenience, we use $\alpha=\{n,k\}$ to denote both the band
index and the Bloch wave vector. Not considering the interparticle
interaction, every particle in $x\in[0,L/\omega]$ is governed by
\begin{equation}\label{eqn11}
\left[-\frac{\hbar^2}{2m}\frac{\partial^2}{\partial x^2}+\mu
B\cos(2\omega x)\right]\varphi_{\alpha}(x)=E_{\alpha}\varphi_{\alpha}(x).
\end{equation}
Substitute $z=\omega x$, $q=\dfrac{m\mu B}{\hbar^2\omega^2}$ and
$\lambda=\dfrac{2mE}{\hbar^2\omega^2}$ into equation (\ref{eqn11}),
then this equation becomes
\begin{equation}\label{eqn12}
\frac{d^2\varphi(z)}{d z^2}+[\lambda-2q\cos(2z)]\varphi(z)=0,
\end{equation}
\begin{equation*}
z\in[0,L].
\end{equation*}

It is called Mathieu's Differential Equation, which can be solved with Randall B. Shirts's \cite{lab22} method.
Using Bloch's theorem, we first set
\begin{align}
\varphi(z)&=\exp(i\nu z)u(z),\\
\varphi(z)&=\varphi(z+\pi),\\
u(z)&=u(z+L),\\
\nu&=\frac{2l}{M},\;\;\left(l\in0,\pm1,\cdots,\pm\frac{M-1}{2}\right).
\end{align}
Since $u(z)$ is periodic with period $\pi$, it can be expanded in
Fourier series:
\begin{equation}\label{eqn17}
\varphi(z)=\exp(i\nu z)\sum_nc_{_n}\exp(i2nz).
\end{equation}

Then substitute (\ref{eqn17}) into (\ref{eqn12}), we obtain an infinite
symmetric tridiagonal matrix equation, that is, eigen equation.
\begin{multline}
\begin{pmatrix}
\ddots & \cdots & \cdots & \cdots & \cdots & \cdots & \cdots \\
\cdots & (\nu-4)^2 & q & 0 & 0 & 0 & \cdots\\
\cdots & q & (\nu-2)^2 & q & 0 & 0 & \cdots \\
\cdots & 0 & q & \nu^2 & q & 0 & \cdots\\
\cdots & 0 & 0 & q & (\nu+2)^2 & q & \cdots\\
\cdots & 0 & 0 & 0 & q & (\nu+4)^2 & \cdots\\
\cdots & \cdots & \cdots & \cdots & \cdots & \cdots & \ddots
\end{pmatrix}\\
\times\begin{pmatrix} \vdots \\ c_{_{-2}} \\ c_{_{-1}} \\ c_{_0} \\
c_{_1} \\ c_{_2} \\\vdots
\end{pmatrix}
=\lambda
\begin{pmatrix}
\vdots \\ c_{_{-2}} \\ c_{_{-1}} \\ c_{_0} \\ c_{_1} \\ c_{_2}
\\\vdots
\end{pmatrix}.
\end{multline}

By truncating the matrix in each direction (centered at the smallest
diagonal element $\nu^2$) at sufficiently large dimensions,
approximations to the desired eigenvalues can be obtained to any
desired precision. We could obtain the eigenvalue and eigenfunction
easily using Mathematica or Matlab. In practice, the eigenvalue and
eigenfunction converge quickly as the dimension increases when $q$
is not too large. For example, if we choose $\nu=6/7$ and $q=1$, the
relative difference of the lowest eigenvalue between truncating the
matrix at $21$-D and $2001$-D is only $10^{-38}$. For large values
of $q$ and for high orders, we can use
asymptotic expansions instead \cite{lab18,lab19}. since the matrix dimension that needed to get
accurate eigenvalues and eigenfunctions becomes large.

With Girardeau's Fermi-Bose mapping (\ref{eqn7}), the eigenstates of
the hard-core Bose system is given by the Slater determinant
\begin{multline}
\psi^B(x_{_1},\cdots,x_{_N})=\frac{A(x_{_1},\cdots,x_{_N})}{\sqrt{N!}}
\\\times
\begin{vmatrix}\varphi_{\alpha_{_1}}(x_{_1})& \varphi_{\alpha_{_2}}(x_{_1})&
\cdots&
\varphi_{\alpha_{_N}}(x_{_1})\\\varphi_{\alpha_{_1}}(x_{_2})&
\varphi_{\alpha_{_2}}(x_{_2})& \cdots&
\varphi_{\alpha_{_N}}(x_{_2})\\\vdots& \vdots& \vdots&
\vdots\\\varphi_{\alpha_{_1}}(x_{_N})&
\varphi_{\alpha_{_2}}(x_{_N})& \cdots& \varphi_{\alpha_{_N}}(x_{_N})
\end{vmatrix}%
\quad.
\end{multline}
The total energy of the system is
\begin{equation}
E=\sum^N_{i=1}E_{\alpha{_i}}.
\end{equation}
As a result of the Bose-Fermi mapping, the energy spectrum of the
Bose and corresponding Fermi system are identical. When the
temperature is 0, the system will be on the ground state
--- $N$ particles on the $N$ lowest eigenstates, respectively.

Calculations show that the energy gap $\triangle E$ between the first and the
second Bloch bands decreases with the decrease of $M$, and when $M$ is large, the energy gap $\triangle E$ is independent of $M$. The energy gap between the first and the
second Bloch bands as a function of magnetic field $B$ and frequency
$\omega$ are plotted in Fig. 1, respectively. At $B=0$, raised by
Girardeau \cite{lab2}, the energy gap of TG gas is the smallest (not
0). The energy gap increases as $B$ and $\omega$ increase. As the
Boltzman Distribution Law notes, $\dfrac{n_{_k}}{n_{_{k'}}}\propto
\exp(-\dfrac{\triangle E}{kT})$, with $\triangle E=E_k-E_{k'}$, ${n_{_k}}$ and $n_{_{k'}}$ are particle numbers on the corresponding states. So the system have a bigger probability in the ground state with large $B$ and $\omega$
than the case $B=0$ when the total number $N=M$.
That is to say, one could realize ground-stated TG system more simple
with large $q$ and $\omega$ than $q=0$ when the total number $N=M$.

\begin{figure}[h]
\includegraphics[scale=0.7]{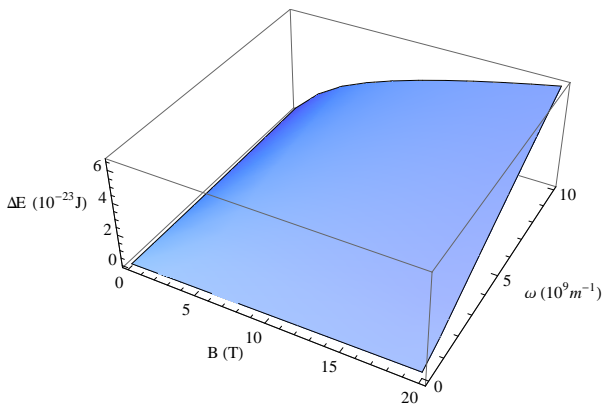}
\caption{Plots of the energy gap between the first and the second
Bloch bands as a function of magnetic induction $B$ and frequency
$\omega$. $M=9$, $m=1.44\times10^{-25}\rm kg$,
$\mu=9.274\times10^{-24}\rm J\cdot \rm T^{-1}$.}
\end{figure}

\section{properties of the TG gas compared with the fermi gas}\label{sec:level3}

In this section, we further study properties of the 1D $N$ magnetized
hard-core Bosons in the external magnetic field (\ref{eqn10}), and then
compared with the properties of 1D $N$ magnetized Fermions in the
``same'' state which hard-core Bosons occupy.

\subsection{Reduced single-particle density matrix and Single-particle density}\label{subsec:level1}
The reduced single-particle density matrix with normalization $\int
\rho(x,x)\, \mathrm{d}x=N$ is given by
\begin{multline}\label{eqn25}
\rho^B(x,x',t)\equiv N\int
{\psi^B}^*(x,x_{_2},\cdots,x_{_N},t)\\
\times\psi^B(x',x_{_2},\cdots,x_{_N},t)\,\mathrm{d}x_{_2}\cdots
\,\mathrm{d}x_{_N}.
\end{multline}

For the Fermi gas, the reduced single-particle density matrix can
be simplified to
\begin{equation}\label{eqn26}
\begin{split}
\rho^F(x,x',t)&=N\int {\psi^F}^*(x,x_{_2},\cdots,x_{_N},t)\\
&\times\psi^F(x',x_{_2},\cdots,x_{_N},t)\,\mathrm{d}x_{_2}\cdots\,\mathrm{d}x_{_N}\\
&=\sum_{\substack{\alpha=all\,
states\\occupyed}}{\varphi_{_\alpha}(x,t)}^*\varphi_{_\alpha}(x',t).
\end{split}
\end{equation}

\begin{figure}[h]
\includegraphics[scale=0.8]{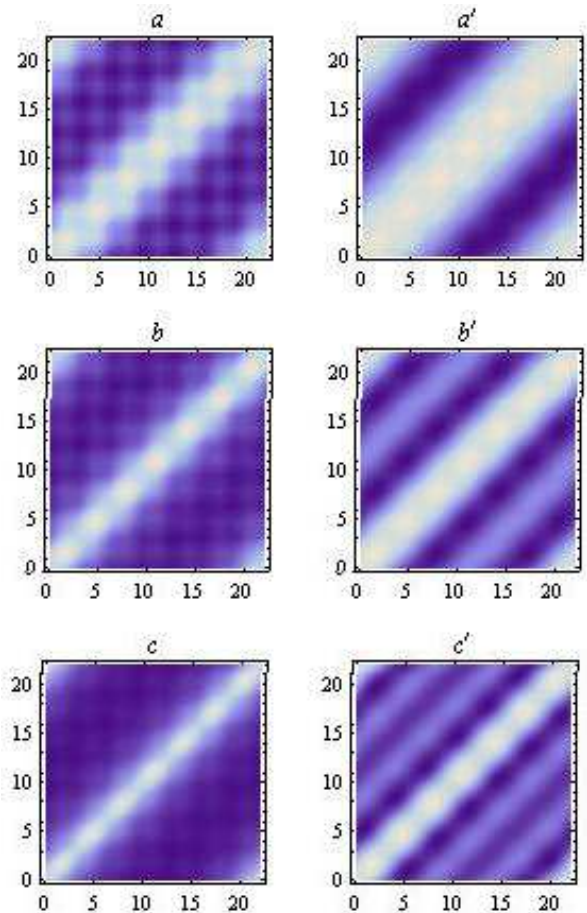}
\caption{Density-plots of $\rho(z,z')$ of the hard-core TG gas
($a$,$b$,$c$) and Fermi gas ($a'$,$b'$,$c'$), $M=7$. Abscissa is
$z_{_1}$, Ordinate is $z_{_2}$. ($a$,$a'$) $N=3$; ($b$,$b'$) $N=5$;
($c$,$c'$) $N=7$. }
\end{figure}

Both the reduced single-particle matrices of hard-core bosons and
fermions in the ground state in the external magnetized field
(\ref{eqn10}) are plotted separately in Fig. 2. The multidimensional
integral in equation (\ref{eqn25}) is evaluated numerically by Monte
Carlo integration using Mathematica. Because the relationship between
$x$ and $z$ is proportional, for the sake of simplicity, we take $z$
as a variable. From the figure we can see the off-diagonal elements
of the matrix of both the two kinds of gases decay quickly as $N$ increases;
The bright diagonal stripe of TG gas is thinner than Fermi gas under
the same condition. This reflects the condensate properties of Bose
gas; And the shock of the off-diagonal elements $\rho(z,M\pi-z)$ of
TG gas is smaller than Fermi gas. To see this more clearly, we have
plotted the off-diagonal elements $\rho(z,M\pi-z)$ in Fig. 3. Sudden
rise in the border is the result of using periodic boundary
conditions.

\begin{figure}[h]
\includegraphics[scale=1]{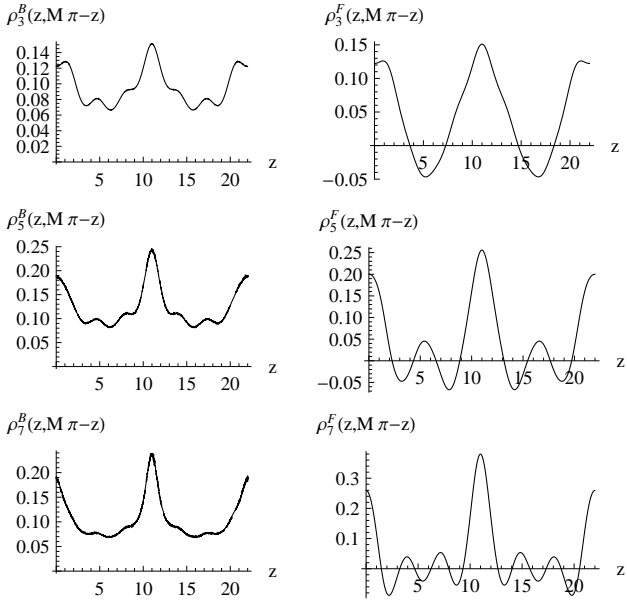}
\caption{Plots of $\rho(z,M\pi-z)$, $M=7$. $\rho^B_n(z,M\pi-z)$
stands for $\rho(z,M\pi-z)$ of $n$ hard-core Boson particles, so as
$\rho^F_n(z,M\pi-z)$}.
\end{figure}

 Using the mapping (\ref{eqn7}), we can
prove that the single-particle density $\rho(x,t)$, normalized to
$N$, of both hard-core Bosons and Fermions are equal, no matter
whether they are on the ground-state or not.
\begin{equation}\label{eq:barwq}
\begin{split}
\rho^B(x,t)&=N\int
|\psi^B(x,x_{_2},\cdots,x_{_N},t)|^2\,\mathrm{d}x_{_2}\cdots
\,\mathrm{d}x_{_N}\\
&=N\int|\psi^F(x,x_{_2},\cdots,x_{_N},t)|^2\,\mathrm{d}x_{_2}\cdots
\,\mathrm{d}x_{_N}\\
&=\sum_{\substack{\alpha=all\,states\\occupyed}}|\varphi_{_\alpha}(x,t)|^2\\
&=\rho^F(x,t)
\end{split}
\end{equation}
 $\rho(x,t)$ is just the
element of the reduced single-particle matrix $\rho(x,x',t)$ when
$x=x'$. Fig. 2 shows that $\rho(x)$ is cyclical, which indicates
that particles tend to stay cycle at where its potential energy is low.

\subsection{Pair distribution function}\label{subsec:level2}

The pair distribution function, which is the probability of finding
a second particle as a function of distance from an initial
particle, normalized to $N(N-1)$, is defined as
\begin{equation}
\begin{split}
&D(x_{_1},x_{_2},t)\\&=N(N-1)\int
|\psi^B(x_{_1},x_{_2},\cdots,x_{_N},t)|^2\,\mathrm{d}x_{_3}\cdots
\,\mathrm{d}x_{_N}\\
&=N(N-1)\int
|\psi^F(x_{_1},x_{_2},\cdots,x_{_N},t)|^2\,\mathrm{d}x_{_3}\cdots
\,\mathrm{d}x_{_N}=\\
&\frac{1}{2}\sum_{\substack{\alpha,\alpha'=all\, states\\ occupied}}
|\varphi_{_\alpha}(x_{_1},t)\varphi_{_{\alpha'}}(x_{_2},t)-\varphi_{_\alpha}(x_{_2},t)\varphi_{_{\alpha'}}(x_{_1},t)|^2.
\end{split}
\end{equation}
From the above equation we can see that the pair distribution
functions of both hard-core Boson gas and Fermi gas are identical.
In fact, the average of particle's coordinate and potential of both
the two gases are identical, too. Because they involve absolute
values of the wave functions only. Also we can say this feature is
determined by the fact that the single particle density of both
hard-core Boson gas and Fermi gas are exactly of the same. For the
ground state of a system with $N$ particles governed by Hamilton
(\ref{eqn9}), the pair distribution function is
\begin{equation}
D(x_{_1},x_{_2})=\frac{1}{2}\sum_{\alpha,\alpha'=0}^{N-1}|\varphi_{_\alpha}(x_{_1})\varphi_{_{\alpha'}}(x_{_2})-\varphi_{_\alpha}(x_{_2})\varphi_{_{\alpha'}}(x_{_1})|^2.
\end{equation}

Figure 4 shows density-plots of the pair distribution function of
the ground state for different number of particles. when
$x_{_1}=x_{_2}$, $D(x_{_1},x_{_2})=0$. It is the result of the
impenetrable hard-core interparticle interaction. Just like the
single particle density, the pair distribution function reflects the
same periodicity and particles tending to tarry at the low potential
energy. As the number of particle increases, the black strip becomes
thinner. One of the reason is that the averaged distance between
particles reduces as the number of particles increases while the
scope of the potential field is certain. In two distant regions,
away from the diagonal in Fig. 4, $D(z_{_1},z_{_2})$ is cyclical.
This reflects the hard-core interaction between particles is local.

\begin{figure}[h]
\includegraphics[scale=0.65]{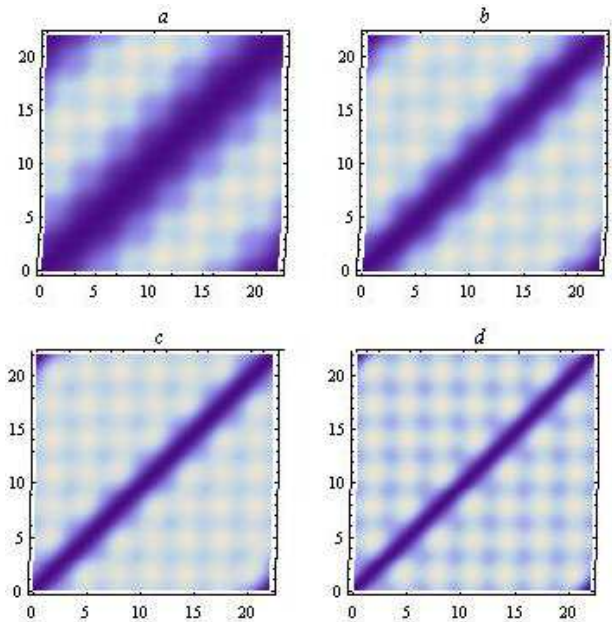}
\caption{Density-plots of the pair distribution function
$D(z_{_1},z_{_2})$. Abscissa is $z_{_1}$, Ordinate is $z_{_2}$.
$M=7$. (a) $N=2$; (b) $N=3$; (c) $N=5$; (d) $N=7$.}
\end{figure}

\subsection{Momentum distribution}\label{subsec:level3}
The momentum distribution, related to the reduced density matrix, is
defined as
\begin{equation}
n(k)=\frac{1}{2\pi}\int
\rho(x,x')e^{-ik(x-x')}\,\mathrm{d}x\,\mathrm{d}x',
\end{equation}
with the normalization
\begin{equation*}
\int n(k)\,\mathrm{d}k=N.
\end{equation*}
Actually, the momentum distribution is just the Fourier
transformation of the reduced density matrix (\ref{eqn25}). We know
that the Fourier transform of the original function and the
transformed function is one-to-one, as the difference of the reduced
density matrix $\rho(x,x')$ of the hard-core Bose gas and Fermion
gas, the momentum distribution is surely not the same. Yuan Lin and
Biao Wu \cite{lab9} have plotted the momentum distributions for both
the TG gas and the free Fermi gas in a periodic Kronig-Penney
potential, which shows the difference --- Even a free Fermi gas has
a broader momentum distribution than the most strongly interacting
boson gas. We would like to add that, although their momentum
distribution functions are different, their total momentum are of
the same. With the equation
\begin{equation}
(\frac{\partial}{\partial x_{_1}}+\frac{\partial}{\partial
x_{_2}})\mathrm{sgn}(x_{_2}-x_{_1})=0,
\end{equation}
we can prove
\begin{equation}
\sum_{i=1}^N\frac{\partial}{\partial
x_{_i}}A(x_{_1},\cdots,x_{_N})=0.
\end{equation}
And then, go ahead, we can obtain
\begin{equation}
\bar{P}(t)^B=\bar{P}(t)^F,\,\,\mathrm{where}\,\,
P=-i\hbar\sum_{i=1}^N\frac{\partial}{\partial x_{_i}}.
\end{equation}

\section{summary and conclusions}

On solving the eigenfunction of the TG gas in the external magnetic field, we found that the TG system with large $B$ and $\omega$ has a bigger probability in the ground state than the TG gas raised by Girardeau in 1960 when the number of bosons $N$ is commensurate with the number of potential cycles $M$. It reveals that we can realize ground-stated TG system more easier. And then we have studied properties of magnetized TG gas and
Fermi gas in an external magnetic field. comparing with each other when they are in the ``same'' state reveals that it is impossible to distinguish them just from their average value of particle's coordinate and potential energy, system's otal momentum, single-particle density and pair distribution function. But we could distinguish them from their reduced single-particle matrices or their momentum distributions. In order to illustrate our results, we have plotted their reduced single-particle matrices and the momentum distributions when they are in the ground state. Although their momentum distribution functions are different, their total momentum are of the same. These results will be deeply helpful for understanding the TG gas.

\section*{Acknowledgment}

This work has been supported by the National Natural Science Foundation of China under Grant No. 10975125.

\bibliography{apssamp}

\end{document}